\documentclass[showpacs,preprintnumbers,amsmath,amssymb]{revtex4}
\usepackage{amsmath,amssymb,graphics,epsfig,subfigure}
\usepackage{color}

\begin{document}
\newcommand {\nn} {\nonumber}
\renewcommand{\baselinestretch}{1.3}

\title{Spherical Orbital Dynamics and Relativistic Precession in Kerr-MOG Spacetime}
	
\author{Hui-Min Wang\footnote{wanghuimin@whu.edu.cn}}

\affiliation{School of Physics and Technology, Wuhan University, Wuhan 430072, China}

\begin{abstract}
	We study the dynamics and relativistic precessions of massive particles on spherical orbits around Kerr-MOG black holes in scalar-tensor-vector gravity (STVG). By employing the Hamilton-Jacobi formalism, we derive conserved quantities and analyze how the MOG parameter $\alpha$ and orbital tilt angle $\zeta$ influence the innermost stable spherical orbits (ISSOs) and orbital stability. We compute the nodal and periastron precession frequencies, finding that nodal precession increases monotonically with both black hole spin and MOG parameter, while periastron precession exhibits a more complex behavior: MOG amplifies curvature-induced effects, which can be partially counteracted by spin. Furthermore, to complement the orbital analysis, we examine the Lense-Thirring spin precession of a gyroscope and demonstrate its sensitivity to the MOG parameter, spin, and orbital tilt angle. These results reveal distinctive signatures of modified gravity in orbital dynamics and provide a potential observational probe to test deviations from general relativity near rotating black holes.
\end{abstract}

%\pacs{04.50.Kd, 04.25.-g, 04.70.-s}
	
\maketitle
	
\section{Introduction}
\label{Introduction}

Black holes provide a unique laboratory for testing gravitational physics in both theoretical and observational contexts. Among them, the Kerr solution, describing a rotating black hole in general relativity (GR), captures essential strong-field effects such as frame dragging, jet formation, and quasi-periodic oscillations (QPOs). Observations of accretion disks and relativistic jets around active galactic nuclei (AGNs) have provided mounting evidence for such phenomena, motivating further exploration of black hole dynamics in realistic astrophysical settings.

Recent developments in alternative theories of gravity have extended the landscape beyond GR. Scalar-tensor-vector gravity (STVG), also known as Modified Gravity (MOG), is a relativistic gravitational theory proposed by Moffat as an alternative to dark matter paradigms \cite{Moffat:2005si}. The theory introduces a massive Proca-type vector field coupled to a scalar field that dynamically determines the effective gravitational coupling strength, $G = G_{\rm N}(1 + \alpha)$, where $\alpha$ is a dimensionless parameter. This framework has demonstrated success in explaining a range of astrophysical phenomena, including galactic rotation curves and cluster dynamics, without invoking dark matter \cite{Moffat:2004bm,Moffat:2013sja,Brownstein:2007sr,Moffat:2007yg}. 

Unlike $f(R)$ gravity, which modifies the Einstein-Hilbert action, or Horndeski theory, which introduces higher-derivative terms, MOG incorporates a massive vector field and a variable gravitational constant. These features lead to an enhanced effective gravitational coupling at galactic and cosmological scales. The theory admits stationary axisymmetric black hole solutions that generalize the Kerr metric through the $\alpha$-dependent modification of the gravitational potential, providing a theoretically consistent framework for testing strong-field gravity deviations from GR.

Building on this framework, a number of recent studies have investigated various aspects of black holes in MOG theory. In particular, the enhanced gravitational coupling has been shown to influence the formation and growth of supermassive black holes \cite{ZhoolidehHaghighi:2023kgs}. Observable signatures of Kerr-MOG black holes have been extensively analyzed, including shadow morphology and light deflection \cite{Wang:2018prk,Moffat:2019uxp,Kuang:2022ojj}, with further generalizations to Kerr-MOG-(anti-)de Sitter (AdS) spacetimes \cite{Liu:2024lbi}. Higher-dimensional extensions of MOG black holes have revealed distinct geometric and thermodynamic properties \cite{Cai:2020igv}, and polarized synchrotron ring images have been studied as possible observational signatures \cite{Qin:2022kaf}. Moreover, the dynamics and emission mechanisms of relativistic jets in Kerr-MOG spacetime have been explored through particle motion simulations and modified Blandford–Znajek magnetospheres, revealing deviations from GR and providing prospects for strong-field tests \cite{LopezArmengol:2016nyi,Camilloni:2023wyn}. Complementary to these developments, recent work in the Kerr-Newman framework has yielded exact analytic solutions for null and timelike geodesics---including spherical, inspiraling, and plunging orbits---using elliptic integrals and Jacobian elliptic functions, offering valuable insight into the role of black hole charge and spin in shaping relativistic trajectories \cite{Wang:2022ouq,Ko:2023igf}. Alongside these efforts, ongoing advances in gravitational wave astronomy provide new opportunities to test MOG theory in the strong-field regime.
 
A central aspect of black hole astrophysics is the study of orbital dynamics, particularly equatorial orbits in Kerr spacetime, which have been widely analyzed. However, in realistic astrophysical settings, orbits are often inclined due to random initial conditions, external perturbations, or interactions with ambient matter. Such spherical orbits---defined by constant radial coordinate but variable latitude---play an important role in modeling accretion disks \cite{Bardeen:1975zz,Petterson1977a,Petterson1977b}. Theoretical studies have investigated the innermost stable spherical orbits (ISSOs) and their dependence on spin and inclination \cite{Kopacek:2024pfd}, revealing their importance in shaping accretion flow structure and emission patterns.

Recent Very Long Baseline Interferometry (VLBI) observations of M87* have identified periodic oscillations in the direction of its relativistic jet, consistent with Lense-Thirring precession caused by a misaligned accretion disk \cite{Cui:2023uyb}. These results support earlier predictions of frame dragging around rotating black holes \cite{Lense:1918zz,Mashhoon:1984fj}, and have sparked renewed interest in relativistic precession phenomena \cite{Chakraborty:2012wv,Chakraborty:2013naa,Wu:2023wld,Pradhan:2024jav,Meng:2024gcg}. In this context, prior studies have explored how orbital motion around rotating black holes encodes information about the final spin of black hole mergers, offering useful guidance for modeling relativistic dynamics \cite{Jai-akson:2017ldo}. Analyzing spin-induced and geometric precessions in modified gravity may help identify potential deviations from GR.

Despite increasing attention to spherical orbits in Kerr spacetime, their counterparts in Kerr-MOG spacetime remain underexplored. The combined complexity of orbital inclination and MOG modifications makes analytical treatment challenging. However, advances in computational techniques now enable precise analysis of such orbits and their associated precessional dynamics. These studies are particularly relevant for future gravitational wave detectors such as LIGO, Virgo, and KAGRA, which may detect subtle signatures of modified gravity encoded in orbital motion and spin precession.

In this work, we conduct a detailed analysis of spherical orbits in Kerr-MOG spacetime, with particular emphasis on how the MOG parameter $\alpha$ and orbital tilt angle $\zeta$ affect the conserved quantities and relativistic precession frequencies. We examine three precessions relevant to strong-field dynamics: nodal precession arising from frame dragging, periastron precession induced by spacetime curvature, and the spin precession of a test gyroscope along the orbit. While previous studies \cite{Pradhan:2020nno,Pradhan:2024jav} have investigated how generalized and Lense-Thirring precession frequencies can distinguish between non-extremal black holes, extremal configurations, and naked singularities in Kerr-MOG spacetime, our focus lies in quantifying how these precessional effects are modulated by $\alpha$ and $\zeta$ across a broad parameter space. This analysis aims to improve our understanding of test particle dynamics in MOG theory and to provide a theoretical basis for identifying potential observational signatures in gravitational wave astronomy and black hole phenomenology.

This paper is structured as follows. In Sec.~\ref{MOGbh}, we review the MOG theory and Kerr-MOG black hole solution. Sec.~\ref{sphorbit} is devoted to analyzing spherical orbits and their dynamical features. In Sec.~\ref{Orbitalprecession}, we compute nodal and periastron precession frequencies and investigate their dependence on orbital parameters and MOG corrections. In Sec.~\ref{LTprecession}, we examine the Lense-Thirring precession of a test gyroscope. We conclude in Sec.~\ref{summary} with a summary and discussion of future directions. Throughout this work, we adopt geometrized units with $G_{\rm N} = c = 1$.

\section{Modified Gravity and Rotating Black Hole Solutions}
\label{MOGbh}

In this section, we provide a concise overview of the core principles of MOG theory and the framework it establishes for analyzing rotating black holes. By extending GR through the inclusion of a scalar field, a vector field, and a variable gravitational constant, MOG provides a framework for describing gravitational phenomena beyond the predictions of GR.

The scalar field in MOG governs the strength of gravitational coupling, allowing it to vary across space and time, in contrast to the fixed coupling constant in standard GR. The vector field introduces a repulsive gravitational component at short distances, which is essential for accounting for galactic rotation curves without the need for dark matter. Furthermore, gravity in MOG theory is nonlinear, meaning its behavior depends not only on the distribution of matter but also on the local intensity of the gravitational field. This theory modifies the Newtonian gravitational constant, allowing it to vary with physical conditions.

To capture these modifications quantitatively, the action of MOG theory, which governs the dynamics of the theory, is articulated by the following formula \cite{Moffat:2005si}:
\begin{eqnarray}
	S=S_{\texttt{GR}}+S_{\phi}+S_{\rm S}+S_{\texttt{M}},
\end{eqnarray}
where the individual terms are given by:
\begin{eqnarray}
	S_{\texttt{GR}}&=&\frac{1}{16\pi}\int d^{4}x\sqrt{-g}\frac{R}{G},\\
	S_{\phi}&=&\int d^{4}x\sqrt{-g}\,\Big(-\frac{1}{4}B^{\mu\nu}B_{\mu\nu}
	+\frac{1}{2}\mu^{2}\phi^{\mu}\phi_{\mu}\Big),\\
	S_{\rm S}&=&\int
	d^4x\sqrt{-g}\biggl(\frac{1}{G^3}\Big(\frac{1}{2}g^{\mu\nu}\nabla_\mu
	G\nabla_\nu G-V(G)\Big)+\frac{1}{\mu^2G}\Big(\frac{1}{2}g^{\mu\nu}\nabla_\mu\mu\nabla_\nu\mu
	-V(\mu)\Big)\biggr).
\end{eqnarray}
Here, $\phi^{\mu}$ is a Proca-type massive vector field with mass $\mu$, and $B_{\mu\nu} = \partial_{\mu}\phi_{\nu} - \partial_{\nu}\phi_{\mu}$ is the corresponding field strength tensor. The functions $V(G)$ and $V(\mu)$ represent the self-interaction potentials for the scalar fields $G(x)$ and $\mu(x)$, respectively. $S_{\texttt{M}}$ denotes the matter action. The mass scale $\mu$ controls the effective range of the vector field, which becomes relevant on kiloparsec scales. Since long-range effects are negligible in black hole solutions, the vector mass $\mu$ can be safely omitted.

In vacuum, $G$ can be treated as a constant, independent of the spacetime coordinates. Under these assumptions, the action simplifies to:
\begin{eqnarray}
	S=\int d^{4}x\sqrt{-g}\left(\frac{R}{16\pi G}-\frac{1}{4}B^{\mu\nu}B_{\mu\nu}\right).
\end{eqnarray}
The tensor field $B_{\mu\nu}$ satisfies the following equations:
\begin{eqnarray}
	\nabla_{\nu}B^{\mu\nu}&=&0,\\
	\nabla_{\sigma}B_{\mu\nu}+\nabla_{\mu}B_{\nu\sigma}+\nabla_{\nu}B_{\sigma\mu}&=&0.
\end{eqnarray}

The relationship between the effective gravitational constant $G$ and the Newtonian one $G_{\rm N}$ is defined as $G = G_{\rm N}(1 + \alpha)$, where $\alpha$ is a dimensionless parameter that quantifies deviations from GR. This enhancement of $G$ reflects the scale-dependent nature of gravity in MOG, offering a mechanism to explain galactic rotation curves, lensing effects, and large-scale structure without invoking dark matter. When $\alpha = 0$, the theory reduces exactly to GR, making $\alpha$ a key parameter for theoretical modeling and observational tests. Astronomical data can be used to constrain its value, and we refer to $\alpha$ as the MOG parameter throughout this work.

In the Boyer-Lindquist coordinate system, the solution for a rotating Kerr-MOG black hole is given by \cite{Moffat:2014aja}:
\begin{equation}
	ds^2=-\frac{\Delta-a^2\sin^2\theta}{\rho^2}dt^2-2a\sin^2\theta\biggl(\frac{r^2+a^2-\Delta}{\rho^2}\biggr)dtd\phi\\
	+\sin^2\theta\biggl(\frac{(r^2+a^2)^2-\Delta a^2\sin^2\theta}{\rho^2}\biggr)d\phi^2+\frac{\rho^2}{\Delta}dr^2+\rho^2d\theta^2,
\end{equation}
where
\begin{equation}
	\Delta=r^2-2GMr+a^2+\alpha G_{\rm N}GM^2,\quad \rho^2=r^2+a^2\cos^2\theta.
\end{equation}
This solution is derived by applying a modified Newman-Janis algorithm to the static MOG metric. This method ensures that the resulting rotating solution remains consistent with the original static spacetime. The Proca-type vector field modifies the gravitational potential, introducing the $\alpha$-dependent term in $\Delta$, which enhances the effective gravitational coupling. 

Solving $\Delta = 0$, we find that, similar to the Kerr black hole, the MOG black hole is characterized by the presence of two horizons: 
\begin{equation}
	r_\pm=G_{\rm N}(1+\alpha)M\biggl(1\pm\sqrt{1-\frac{a^2}{G_{\rm N}^2(1+\alpha)^2M^2}-\frac{\alpha}{1+\alpha}}\biggr).
\end{equation}
Compared to the Kerr case ($\alpha = 0$), the presence of the MOG parameter $\alpha$ shifts the horizons outward, reflecting the enhanced effective gravitational attraction in MOG theory.

\section{Spherical Orbits and Stability in Kerr-MOG Spacetime}
\label{sphorbit}

Building on the framework introduced in Sec.~\ref{MOGbh}, we now examine the motion of massive particles in Kerr-MOG spacetime. In particular, we focus on spherical orbits---trajectories with constant radial coordinate but nontrivial polar motion---as a probe of the three-dimensional structure of curved spacetime and the imprints of modified gravity.

We begin by deriving the geodesic equations using the Hamilton-Jacobi formalism, which naturally yields the conserved quantities governing particle motion. We then analyze spherical orbits with arbitrary tilt angles, examining how the MOG parameter $\alpha$ and the tilt angle $\zeta$ influence the energy, angular momentum, and the Carter constant. Finally, we explore the ISSOs, studying their dependence on spin, inclination, and deviation from GR. These results offer insights into particle dynamics in MOG theory and inform the interpretation of astrophysical phenomena such as disk truncation and jet precession.

\subsection{Geodesic Structure and Conserved Quantities}

Rather than solving the geodesic equations directly, we adopt the Hamilton-Jacobi formalism, which is particularly powerful in GR and its extensions, as it enables separation of variables in highly symmetric spacetimes and naturally leads to conserved quantities.

Derived from the principle of least action, the Hamilton-Jacobi equation governs particle motion in curved spacetime. In stationary, axisymmetric geometries such as Kerr and Kerr-MOG, the equation is separable due to hidden symmetries associated with a Killing tensor. This allows the motion to be decomposed into radial and polar components, each governed by a first-order equation, and introduces an additional constant of motion: the Carter constant.

The Jacobi action $S$ for a test particle takes the form:
\begin{equation}
	S=-Et+L\phi+S_{r}(r)+S_\theta(\theta)+\frac{1}{2}m^2\lambda,\label{Jaction}
\end{equation}
where $E$, $L$ and $m$ are, respectively, the energy, the angular momentum in the direction of the axis of symmetry, and the rest mass of the test particle. Substituting this ansatz into the Hamilton-Jacobi equation yields the following first-order equations:
\begin{eqnarray}
	\rho^{2}\frac{dt}{d\lambda}&=&a(L-aE\sin^2\theta)+\frac{r^2+a^2}{\Delta}\Big((r^2+a^2)E-aL\Big),\\
	\label{phiequation}
	\rho^{2}\frac{d\phi}{d\lambda}&=&\frac{L}{\sin^2\theta}-aE+\frac{a}{\Delta}\Big((r^2+a^2)E-aL\Big),\\
	\label{Requation}
	\rho^{2}\frac{dr}{d\lambda}&=&\pm\sqrt{{\mathcal R}(r)},\\
	\label{thetaequation}
	\rho^{2}\frac{d\theta}{d\lambda}&=&\pm\sqrt{\Theta(\theta)},
\end{eqnarray}
with the radial and polar potentials defined as:
\begin{eqnarray}
	{\mathcal R}(r)&=&\Big((r^2+a^2)E-aL\Big)^2-\Delta\Big(r^2+{\mathcal K}+(L-aE)^2\Big),\\
	\Theta(\theta)&=&{\mathcal K} +\cos^2\theta\biggl(a^2E^2-\frac{L^2}{\sin^2\theta}-a^2\biggr).
\end{eqnarray}

The Carter constant $\mathcal K$, first identified in Ref. \cite{Carter:1968rr}, quantifies the motion in the polar direction and arises from the hidden symmetry encoded in the spacetime's Killing tensor. In equatorial orbits ($\theta = \frac{\pi}{2}$), $\mathcal K$ vanishes, while off-equatorial orbits yield nonzero values that encode the amplitude of $\theta$-oscillations.

In Kerr-MOG spacetime, the enhanced gravitational coupling $G = G_{\rm N}(1 + \alpha)$ modifies the effective potentials $\mathcal R(r)$ and $\Theta(\theta)$ via the $\alpha$-dependent term, thereby shifting the stability conditions for spherical orbits. These modifications, along with potential vector field contributions, lead to deviations in the Carter constant relative to the Kerr case. As a result, $\mathcal K$ serves as a sensitive probe of orbital geometry, capturing the full three-dimensional structure of motion and encoding the dynamical signatures of MOG theory.

\subsection{Spherical Orbits: Dynamical Features and MOG Effects}

We analyze spherical orbits in Kerr-MOG spacetime using the distant-observer framework introduced in Ref.~\cite{Wei:2024cti}, originally developed in the context of GR. Here, we generalize this approach to MOG, where the enhanced gravitational coupling, controlled by the parameter $\alpha$, modifies the structure and stability of spherical orbits in strong-field regimes. In contrast to local treatments of orbital precession~\cite{AlZahrani:2023xix}, this framework is well suited for interpreting astrophysical observables such as jet precession and disk misalignment in sources like M87*.

The combined influence of black hole spin and MOG correction modifies orbital structure beyond the equatorial plane. From Eq.~(\ref{Requation}) that describes the radial motion ($r$-motion) of massive particles, the conditions for spherical orbits with constant radius are:
\begin{equation}
	{\mathcal R}(r)=0, \quad  {\mathcal R}'(r)=0,
\end{equation}
with the prime denoting a radial derivative. We introduce the tilt angle $\zeta$ to characterize orbital inclination, defined such that the particle oscillates between $\theta = \frac{\pi}{2} \pm \zeta$.

The polar $\theta$-motion exhibits periodic behavior, with test particles oscillating between two angular turning points defined by the condition $\Theta(\theta) = 0$ in Eq.~(\ref{thetaequation}). At these turning points, the Carter constant takes distinct forms. For massive particles:
\begin{equation}
    {\mathcal K} = a^2\sin^2\zeta (1-E^2)+L^2\tan^2\zeta,	\label{massivecarter}
\end{equation}
whereas for photons:
\begin{equation}
	{\mathcal K} = -a^2 E^2\sin^2\zeta+L^2\tan^2\zeta.	\label{photoncarter}
\end{equation}
A detailed analytical derivation of the angular momentum and Carter constant for photon orbits in Kerr-MOG spacetime can be found in Ref.~\cite{BenAchour:2025uzp}, offering a complementary perspective to our treatment of massive particles. While $\alpha$ does not appear explicitly in these expressions, it indirectly affects $\mathcal K$ through its influence on $E$ and $L$.

\begin{figure}
	\begin{center}
		\subfigure[]{\label{rEnergy1a}
			\includegraphics[width=0.31\textwidth]{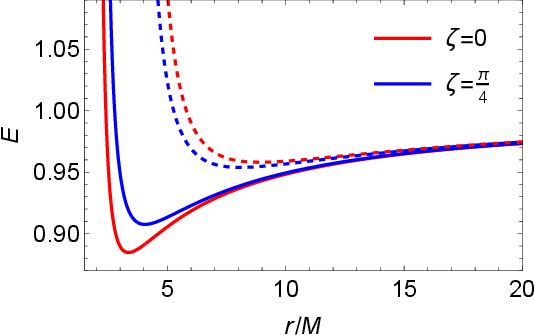}}
		\quad
		\subfigure[]{\label{rMomentum1b}
			\includegraphics[width=0.3\textwidth]{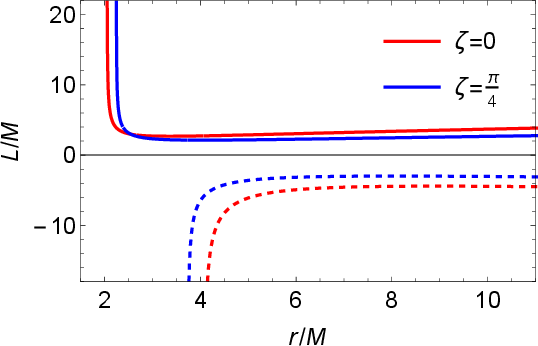}}
		\quad
		\subfigure[]{\label{rCarter1c}
			\includegraphics[width=0.3\textwidth]{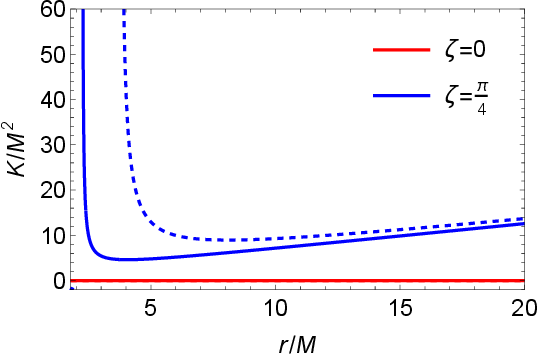}}\\
		\subfigure[]{
			\includegraphics[width=0.31\textwidth]{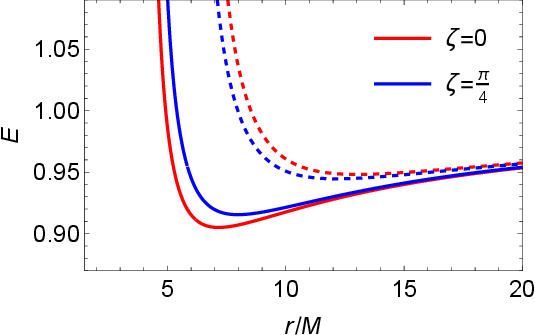}}
		\quad
		\subfigure[]{
			\includegraphics[width=0.3\textwidth]{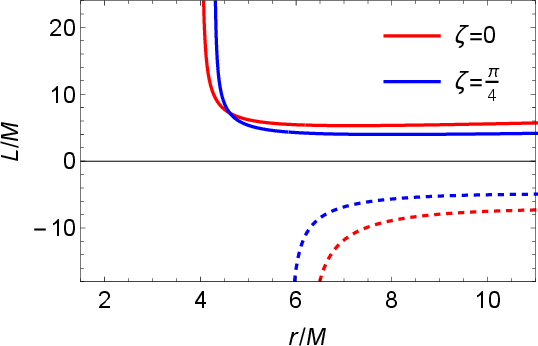}}
		\quad
		\subfigure[]{
			\includegraphics[width=0.3\textwidth]{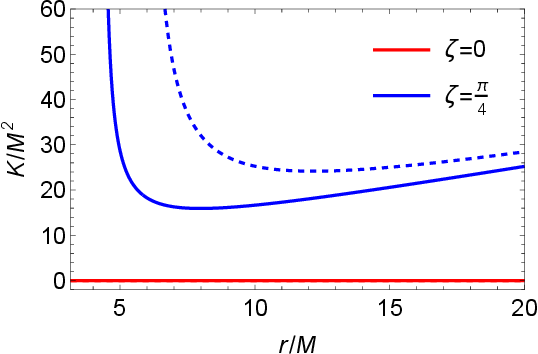}}
		\caption{Energy $E$, angular momentum $L$, and Carter constant ${\mathcal K}$ for spherical orbits as functions of orbital radius $r$, with spin $a = 0.8M$. Top row: $\alpha = 0.1$; bottom row: $\alpha = 1$. Solid/dashed curves represent prograde/retrograde orbits, and red/blue denote $\zeta = 0, \frac{\pi}{4}$.}\label{conserved}
	\end{center}
\end{figure}

We show the orbital-radius dependence of energy, angular momentum, and the Carter constant for varying MOG parameter $\alpha$ and tilt angle $\zeta$ in Fig.~\ref{conserved}. As shown in Fig.~\ref{rEnergy1a}, the particle energy decreases rapidly with orbital radius, reaches a minimum, and then asymptotically approaches a constant value. It is largely insensitive to tilt angle and orbital direction. Numerical checks confirm that at $r = 50M$, the energy values converge to $\sim 0.989$, and at $r = 500M$, they all approach 0.9989, indicating that particle dynamics at large radii are effectively governed by flat spacetime. The energy minimum marks the ISSO: orbits are unstable for $r < r_{\rm ISSO}$ and stable beyond. At fixed radius, prograde orbits exhibit increasing energy with tilt angle, while retrograde orbits show the opposite behavior, consistent across both stable and unstable regimes.

Fig.~\ref{rMomentum1b} shows that the angular momentum $L$ decreases with radius for both prograde and retrograde orbits, eventually flattening at large distances. For stable orbits, $L$ decreases with tilt angle, reaching a maximum in the equatorial plane. As shown in Fig.~\ref{rCarter1c}, the Carter constant $\mathcal K$ vanishes for equatorial orbits and exhibits a non-monotonic dependence on $r$ in spherical motion: it decreases initially and then increases with radius. Retrograde orbits consistently yield larger $\mathcal K$ than prograde ones at the same radius. Since $\mathcal K$ depends on both $E$ and $L$, it serves as a crucial diagnostic for non-equatorial dynamics.

Comparing $\alpha = 0.1$ (top row) and $\alpha = 1$ (bottom row) in Fig.~\ref{conserved}, we find that increasing $\alpha$ shifts the ISSO, energy, angular momentum, and the Carter constant to larger radii. This trend persists across different tilt angles. For instance, for $\zeta = \frac{\pi}{4}$ prograde motion, the ISSO shifts from $r = 4.05M$ ($\alpha=0.1$) to $7.98M$ ($\alpha=1$), with $E$ increasing from 0.9075 to 0.9155. The enhanced gravity shifts stable spherical orbits to larger radii by providing greater centripetal force and requires higher particle energy to maintain equilibrium. Consequently, both the ISSO radius and the associated energy increase with $\alpha$. At a significant distance of $r = 50M$, larger $\alpha$ lowers the energy ($E = 0.9806$ vs. $0.9892$) but increases both angular momentum ($L = 7.40M$ vs. $5.39M$) and Carter constant ($\mathcal{K} = 54.82M^2$ vs. $29.09M^2$) compared to the near-GR case.

At fixed values of $E$, $L$, or $\mathcal{K}$, retrograde orbits consistently lie at larger radii than their prograde counterparts due to frame dragging. Spin-induced dragging lowers the effective potential for prograde motion while raising it for retrograde paths. As a result, retrograde particles require more energy and angular momentum to remain bound, leading to larger orbital radii. This asymmetry diminishes at large distances where both spin and gravity become weak.

\begin{figure}
	\begin{center}
		\subfigure[]{\label{zetaCarter2a}
			\includegraphics[width=0.29\textwidth]{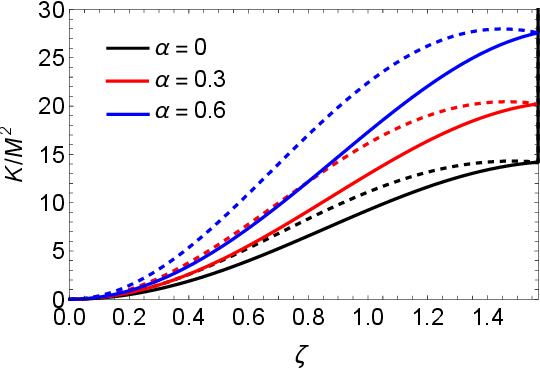}}
		\quad
		\subfigure[]{\label{zetaCarter2b}
			\includegraphics[width=0.31\textwidth]{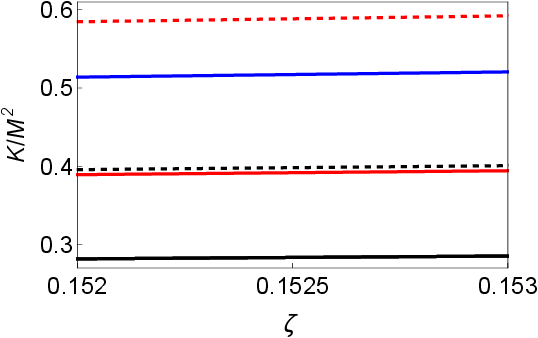}}
		\quad
		\subfigure[]{\label{zetaCarter2c}
			\includegraphics[width=0.31\textwidth]{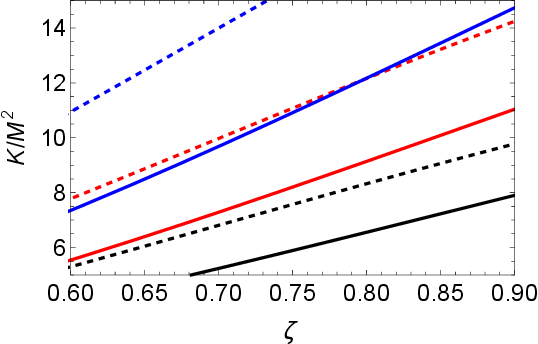}}
		\caption{Carter constant ${\mathcal K}$ as a function of orbital tilt angle $\zeta$ with spin $a = 0.8M$ and orbital radius $r = 10M$. Panels (b) and (c) magnify regions near $\zeta = 0.15$ and $0.7$, respectively, highlighting the $\alpha$-dependence. Solid and dashed curves represent prograde and retrograde motion.}\label{zetaCarter}
	\end{center}
\end{figure}

Fig.~\ref{zetaCarter2a} shows the dependence of $\mathcal K$ on tilt angle $\zeta$ for different $\alpha$. $\mathcal K$ vanishes in the equatorial plane and increases nonlinearly with $\zeta$. At $\zeta = 0$ and $\frac{\pi}{2}$, prograde and retrograde values coincide; otherwise, retrograde orbits yield larger $\mathcal K$. In general, $\mathcal K$ increases with $\alpha$, indicating enhanced non-equatorial dynamics in MOG. However, as Fig.~\ref{zetaCarter2b} shows, at small angles ($\zeta \sim 0.15$), retrograde Kerr orbits can have higher $\mathcal K$ than prograde MOG orbits, suggesting that $\mathcal K$ alone cannot distinguish between the theories in this regime. In contrast, Fig.~\ref{zetaCarter2c} demonstrates that for $\zeta \gtrsim 0.8$, $\mathcal K$ increases monotonically with $\alpha$ across all orbital directions, making it a reliable diagnostic for detecting MOG effects at larger inclinations.

\subsection{Stability and Parameter Dependence of ISSOs}

Among spherical orbits, the ISSO plays a central role in the physics of black hole accretion. It defines the inner edge of the disk, separating stable from dynamically unstable configurations. Inside the ISSO, particles undergo rapid inspiral; outside, they can support long-lived orbits and the formation of a disk. The ISSO location thus encodes important information about the black hole spacetime. Stability requires that the radial effective potential exhibits a local minimum, yielding the condition:
\begin{equation}
	{\mathcal R}''(r)=0.
\end{equation}

We compute the ISSO radius $r_{\rm ISSO}$, energy $E$, angular momentum $L$, and Carter constant $\mathcal K$ for prograde and retrograde orbits under varying the MOG parameter $\alpha$ and tilt angle $\zeta$, summarized in Table~\ref{tableisco}. For prograde motion, $r_{\rm ISSO}$ and $E$ increase with $\zeta$, while retrograde orbits show the opposite trend. Angular momentum $L$ decreases with $\zeta$ and vanishes at $\zeta = \frac{\pi}{2}$. Meanwhile, the Carter constant $\mathcal K$ increases monotonically with $\zeta$. Notably, the trends in $L$ and $\mathcal K$ are independent of orbital direction. At fixed $\zeta$, increasing the MOG parameter $\alpha$ shifts $r_{\rm ISSO}$, $L$, and $\mathcal K$ to larger values for both orbital directions. The behavior of $E$ exhibits more complex dependence on $\alpha$: for prograde orbits, $E$ increases with $\alpha$ at $\zeta = 0$ and $\frac{\pi}{4}$, but slightly decreases at $\zeta = \frac{\pi}{2}$. For retrograde motion, $E$ always decreases with $\alpha$. These features may reflect deeper structure in the MOG potential. Notably, at $\zeta = \frac{\pi}{2}$, the ISSO radii coincide for both orbital directions, reflecting the symmetry of polar orbits.
\begin{table}[h]
	\begin{center}
		\begin{tabular}{|c|c|cccc|cccc|}
			\hline
			$\quad\alpha\quad$ & $~\quad\zeta\quad~$ &  $\quad r_{\rm ISSO}(p)\quad$  & $\quad E(p)\quad$ & $\quad\quad L(p)\quad\quad$& $\quad{\mathcal K}(p)\quad$&  $\quad r_{\rm ISSO}(r)\quad$  & $\quad E(r)\quad$ & $\quad\quad L(r)\quad\quad$ & $\quad{\mathcal K}(r)\quad$
			\\ 
			\hline
			0 & 0 & 2.9066 & 0.8778  & 2.3804 &0& 8.4318 & 0.9597 & -4.1004 &0 
			\\
			0 & $\pi/4$ & 3.5813 &0.9048 & 1.91857 &3.7389 & 7.6035 & 0.9554 & -2.7652 & 7.6745
			\\
			0 & $\pi/2$ & 5.5680 &0.9396 & 0 & 11.5373 &5.5680 & 0.9396 & 0  & 11.5373 
			\\
			\hline
			1 & 0 & 7.1617 &0.9051 & 5.3075 & 0 & 13.0221 & 0.9482 & -7.1770 &  0 \\
			1 & $\pi/4$ & 7.9775 & 0.9154 &3.9837 & 15.9218 &12.1573 & 0.9446 & -4.9121 & 24.1637  \\ 
			1 & $\pi/2$ & 10.0540 & 0.9332 &0 & 40.2583&10.0540 & 0.9332 & 0 & 40.2583  \\ 
			\hline
		\end{tabular}
		\caption{Spherical orbital radius, energy, angular momentum, and the Carter constant for massive particles at the ISSO. Subscripts $p$ and $r$ denote prograde and retrograde motion, respectively. Spin parameter is fixed at $a = 0.8M$.}\label{tableisco}
	\end{center}
\end{table}

ISSO locations provide a sensitive diagnostic of modified gravity. In MOG, the enhanced coupling $G = G_{\rm N}(1 + \alpha)$ deepens the gravitational potential, pushing ISSOs outward. Meanwhile, the orbital tilt angle $\zeta$ exposes test particles to off-equatorial curvature and frame dragging effects, increasing the Carter constant $\mathcal K$ and modifying the distribution of angular momentum. As $\zeta \to \frac{\pi}{2}$, the orbit becomes polar with vanishing $L$. The Carter constant, which encodes hidden symmetries, grows significantly with both $\zeta$ and $\alpha$, especially at high inclinations. It thus serves as a geometric tracer of both black hole spin and MOG-induced deviations, indicating that spherical orbits in MOG possess greater structural complexity than their GR counterparts.

These dynamical effects may carry observational implications. In systems exhibiting misaligned accretion disks or jet precession---such as M87*---the combined influence of $\alpha$ and $\zeta$ could produce deviations from GR predictions, such as shifts in disk truncation radii or jet precession frequencies. Detecting such signatures through VLBI techniques, e.g., with the Event Horizon Telescope (EHT), could offer direct evidence for modified gravity. Motivated by this possibility, we now turn to a detailed investigation of relativistic precession phenomena---namely, nodal and periastron precession---arising from spin and curvature effects in the Kerr-MOG geometry.

\section{Orbital Precession in Kerr-MOG Spacetime}
\label{Orbitalprecession}

Beyond circular equatorial motion and spherical motion, test particles around rotating black holes experience relativistic precessions arising from frame dragging and spacetime curvature. These effects become significant when orbits are inclined or slightly eccentric. Two primary forms of precession are of interest: nodal precession, the slow rotation of the orbital plane caused by spacetime rotation, and periastron precession, the in-plane advance of the closest approach due to curved geometry. Both phenomena are sensitive to the background spacetime and thus provide robust probes of deviations from GR.

In Kerr-MOG spacetime, the MOG parameter $\alpha$ enhances the gravitational interaction, modifying the effective potential and the rotational structure of the geometry. These deviations from GR directly affect orbital dynamics, particularly in strong-field regions.

A characteristic feature of spacetime rotation is the off-diagonal metric component $g_{t\phi}$, which encodes frame dragging effects. This term introduces coupling between temporal and azimuthal motion, shifting the coordinate angular velocity and modifying the conserved angular momentum. In the azimuthal geodesic equation, $g_{t\phi}$ appears explicitly, and its radial derivative contributes directly to the orbital frequency, especially for tilted orbits near the black hole. In addition, the diagonal components $g_{tt}$, $g_{rr}$, and $g_{\phi\phi}$ determine the spacetime curvature and influence the radial and polar epicyclic frequencies through their derivatives. These govern the deviation from flatness and determine the periastron advance.

When the orbit is tilted, the particle undergoes vertical motion, leading to nodal precession. Small radial deviations cause oscillatory behavior and in-plane precession of the periastron. Both effects are shaped by the spin $a$, the MOG parameter $\alpha$, and the orbital geometry. To quantify these relativistic effects, we derive the fundamental frequencies and define the associated precession rates. We then analyze their dependence on $a$, $\alpha$, and $r$, highlighting their relevance for both particle dynamics and observational signatures.

For circular equatorial orbits, the angular frequency $\Omega_\phi$ is given by
\begin{equation}
\Omega_{\phi}=\frac{-g_{t\phi,r}+\sqrt{(g_{t\phi,r})^2-g_{tt,r} g_{\phi\phi,r}}}{g_{\phi\phi,r}},
\label{omega_phi}
\end{equation}
where a comma denotes partial differentiation. It corresponds to the coordinate angular velocity $d\phi/dt$ in Boyer-Lindquist coordinates. The expression follows from the conserved energy $E$ and angular momentum $L$ associated with the Killing symmetries $\partial_t$ and $\partial_\phi$, together with the normalization condition $u^\mu u_\mu = -1$. The numerator reflects the frame dragging contribution via $g_{t\phi,r}$, while the remaining terms capture the competition between gravitational potential gradient ($g_{tt,r}$) and centrifugal effects ($g_{\phi\phi,r}$). Small perturbations in $r$ and $\theta$ directions induce radial and vertical oscillations, characterized by the radial and vertical epicyclic frequencies, $\Omega_r$ and $\Omega_\theta$, respectively. These frequencies determine the curvature and stability properties, and are key to understanding relativistic precession.

In the remainder of this section, we analyze the two dominant types of orbital precession in Kerr-MOG spacetime. Nodal precession, which is associated with spin-induced frame dragging, arises in tilted orbits and is linked to the frequency difference $\Omega_\phi - \Omega_\theta$. Periastron precession, due to spacetime curvature, occurs even in equatorial motion and is governed by the difference $\Omega_\phi - \Omega_r$. Each type of precession probes a distinct aspect of the underlying geometry and offers potential observables for testing modified gravity theories.

\subsection{Nodal Precession: Frame-Dragging-Induced Orbital Plane Rotation}

Nodal precession results from the Lense-Thirring effect---a manifestation of frame dragging due to black hole spin. This effect causes the orbital angular momentum vector of a particle to precess around the black hole’s spin axis with a frequency proportional to its angular momentum $J = aM$. Mathematically, frame dragging is encoded in the off-diagonal metric component $g_{t\phi}$, which introduces a coupling between time and azimuthal coordinates. In Kerr-MOG spacetime, $g_{t\phi}$ acquires an $\alpha$-dependent correction that enhances this gravitomagnetic interaction, leading to a modified frame dragging rate that scales with both the spin parameter $a$ and the MOG parameter $\alpha$.

The vertical epicyclic frequency $\Omega_\theta$, which governs the out-of-plane motion for tilted orbits, is influenced by frame dragging through the metric component $g_{t\phi}$ and second-order derivatives of the metric with respect to $\theta$ \cite{Ryan:1995wh,Doneva:2014uma}:
\begin{eqnarray}
\Omega_\theta^2 = \frac{1}{2g_{\theta\theta}}\bigg(
X^2 \partial_\theta^2\left(\frac{g_{\phi\phi}}{\mathcal{D}}\right)
- 2XY \partial_\theta^2\left(\frac{g_{t\phi}}{\mathcal{D}}\right)
+ Y^2 \partial_\theta^2\left(\frac{g_{tt}}{\mathcal{D}}\right)
\bigg), \label{omega_theta} 
\end{eqnarray}
where 
\begin{eqnarray}
\mathcal{D} &=& g_{tt}g_{\phi\phi} - g_{t\phi}^2,\\
X &=& g_{tt} + g_{t\phi} \Omega_\phi, \\
Y &=& g_{t\phi} + g_{\phi\phi} \Omega_\phi.
\end{eqnarray}

Nodal precession manifests when the orbital angular frequency $\Omega_\phi$ and the vertical epicyclic frequency $\Omega_\theta$ become detuned due to spacetime rotation, with its characteristic frequency given by \cite{Motta:2013wga}
\begin{eqnarray}
\Omega_{\text{nod}}=\Omega_\phi-\Omega_\theta.
\end{eqnarray}
This frequency quantifies the rate at which the orbital plane precesses around the black hole’s spin axis. Physically, it captures the differential twisting of spacetime experienced by a slightly inclined particle trajectory and provides a direct signature of frame dragging. The asymptotic expansion of $\Omega_{\text{nod}}$ in powers of $1/r$ in the Kerr-MOG geometry yields
\begin{eqnarray}
\Omega_{\text{nod}} \approx 2aM(\alpha +1)\left(\frac{1}{r}\right)^3-\frac{3a^2\sqrt{(\alpha +1) M}}{2}\left(\frac{1}{r}\right)^{7/2}-a M^2\alpha(\alpha +1)\left(\frac{1}{r}\right)^4+\mathcal{O}\left[\frac{1}{r}\right]^{9/2}.
\label{nodal_expansion}
\end{eqnarray}
This expansion highlights the layered contributions from spin and modified gravity: the leading-order $r^{-3}$ term generalizes the Lense–Thirring effect via the enhanced coupling $G = G_{\rm N}(1 + \alpha)$; the $r^{-7/2}$ term originates from higher-order spin–curvature interactions and captures post-Newtonian corrections; and the $r^{-4}$ term, absent in Kerr, arises purely from nonlinear effects unique to MOG. This hierarchy of terms implies that while the $r^{-3}$ scaling dominates at large radii, significant deviations from GR emerge closer to the black hole. 

\begin{figure}
	\center{\subfigure[]{\label{alphanod}
			\includegraphics[width=5cm]{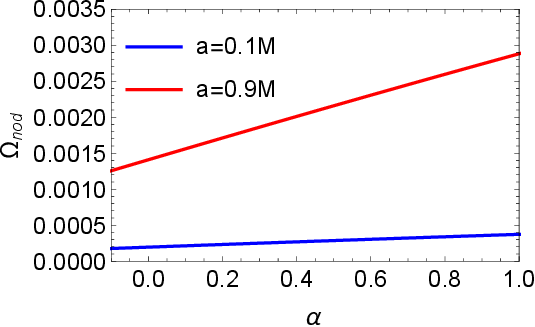}}
		\subfigure[]{\label{anod}
			\includegraphics[width=5cm]{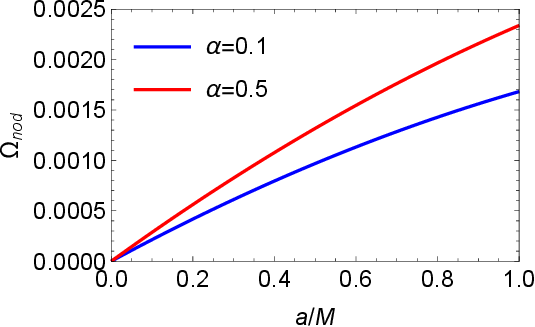}}
		\subfigure[]{\label{rnod}
			\includegraphics[width=5cm]{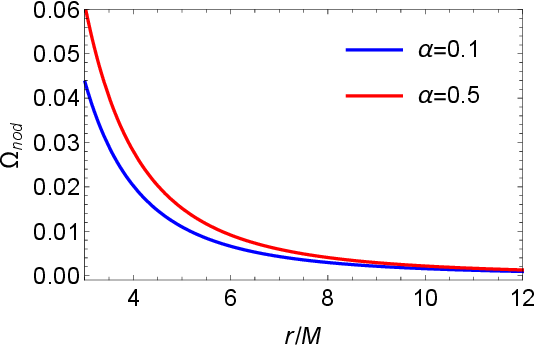}}}
	\caption{Dependence of the nodal precession frequency $\Omega_{\text{nod}}$ on: (a) the MOG parameter $\alpha$ at fixed orbital radius $r=10M$; (b) the spin parameter $a/M$ at $r=10M$; (c) the orbital radius $r/M$ for a rapidly rotating black hole with spin $a=0.9M$.}\label{omeganod}
\end{figure}

Fig.~\ref{omeganod} presents the behavior of the nodal precession frequency $\Omega_{\text{nod}}$ as a function of the MOG parameter $\alpha$, spin parameter $a$, and orbital radius $r$. Fig.~\ref{alphanod} shows that $\Omega_{\text{nod}}$ increases with $\alpha$, reflecting the enhancement of the gravitomagnetic field due to the stronger effective coupling $G = G_{\rm N}(1+\alpha)$ in MOG. This effect is especially pronounced for highly spinning black holes, where frame dragging is already strong. Fig.~\ref{anod} demonstrates the monotonic increase of $\Omega_{\text{nod}}$ with $a$, consistent with the expectation that the Lense-Thirring effect is proportional to the spin angular momentum $J = aM$. Notably, higher values of $\alpha$ amplify this spin dependence, suggesting a nonlinear coupling between spin and modified gravity in the frame dragging sector. Fig.~\ref{rnod} illustrates the steep radial decay of $\Omega_{\text{nod}}$, approximately following an $r^{-3}$ power law at large distances, as predicted by the leading-order term in Eq.~(\ref{nodal_expansion}). The decay is steeper for larger $\alpha$, indicating that MOG-induced deviations from GR are most prominent near the black hole and rapidly suppressed in the weak-field regime.

Taken together, these results confirm that $\Omega_{\text{nod}}$ is enhanced by both the spin and the MOG effects, and suppressed with increasing radius. The sensitivity of nodal precession to strong-field gravity makes it a powerful probe of deviations from GR, with direct implications for jet precession, disk warping, and waveform modulation in EMRIs.

\subsection{Periastron Precession: Curvature-Induced In-Plane Rotation}

We now turn to periastron advance, the second key aspect of relativistic orbital precession, which arises from spacetime curvature rather than frame dragging. It reflects the failure of bound orbits to close in coordinate space due to relativistic corrections and captures the secular in-plane drift of the periapsis over successive revolutions.

Nodal and periastron precession probe complementary aspects of spacetime: the former reflects gravitomagnetic effects due to spin, while the latter encodes curvature via the radial potential structure. As such, simultaneous measurements of both frequencies provide a powerful and complementary tool for testing the gravitational field structure near compact objects and may help distinguish GR from extended theories such as MOG.

Periastron precession describes the advance of the azimuthal location of the periapsis, the point of closest approach within the orbital plane. It is mathematically defined by the mismatch between the azimuthal and radial epicyclic frequencies \cite{Motta:2013wga}:
\begin{equation}
	\Omega_{\text{per}} = \Omega_\phi - \Omega_r,
\end{equation}
where $\Omega_r$ is the radial epicyclic frequency. It is determined from second-order derivatives of the metric components \cite{Ryan:1995wh,Doneva:2014uma}: 
\begin{eqnarray}
	\Omega_r^2 = \frac{1}{2g_{rr}}\bigg(
	X^2 \partial_r^2 \left(\frac{g_{\phi\phi}}{\mathcal{D}}\right)
	- 2XY \partial_r^2 \left(\frac{g_{t\phi}}{\mathcal{D}}\right)
	+ Y^2 \partial_r^2 \left(\frac{g_{tt}}{\mathcal{D}}\right)
   \bigg).
\end{eqnarray}
Since $\Omega_r$ depends on second derivatives of the metric components, its behavior is sensitive to the local spacetime curvature. Specifically, $\Omega_r$ encodes how curvature affects the stability of radial oscillations.

The corresponding expression for $\Omega_{\text{per}}$ in Kerr-MOG spacetime admits the expansion:
\begin{equation}
	\Omega_{\text{per}} \approx 
	\frac{(5\alpha^2 + 11\alpha + 6)M^2}{2\sqrt{(\alpha+1)M}}\left(\frac{1}{r}\right)^{5/2}
	- 4a(\alpha+1)M\left(\frac{1}{r}\right)^{3}
	+ \mathcal{O}(r^{-7/2}).
\label{per_expansion}
\end{equation}
This asymptotic expansion reveals the distinct contributions of spacetime curvature and frame dragging in determining the periastron precession rate. The leading term, proportional to $M^{3/2}r^{-5/2}$ with $\alpha$-dependent enhancement, arises from the radial curvature of the gravitational potential, and reflects the the non-closure of bound orbits. It generalizes the Schwarzschild periastron shift with MOG-induced corrections, which increases the precession rate. The next-order term, scaling as $a r^{-3}$, reflects spin-induced frame dragging and enters with a negative sign, reducing the total precession. While curvature dominates at moderate distances, spin effects become relevant in the near-horizon region, making $\Omega_{\text{per}}$ a useful probe of strong-field gravity and MOG corrections.

\begin{figure}
	\center{\subfigure[]{\label{alphaper}
			\includegraphics[width=5cm]{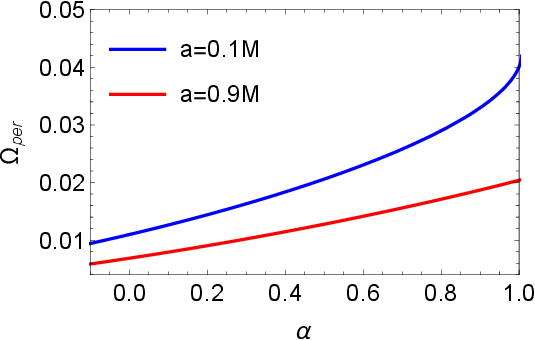}}
		\subfigure[]{\label{aper}
			\includegraphics[width=5cm]{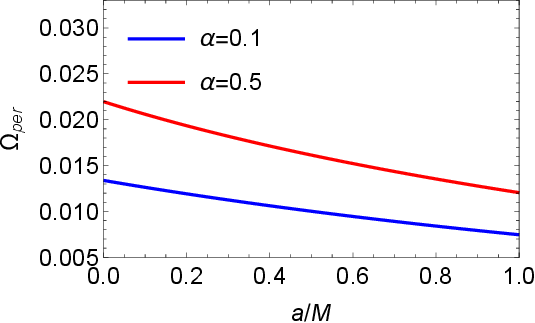}}
		\subfigure[]{\label{rper}
			\includegraphics[width=5cm]{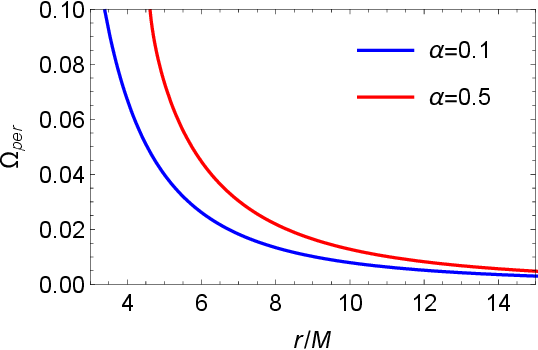}}}
	\caption{Dependence of the periastron precession frequency $\Omega_{\text{per}}$ on: (a) the MOG parameter $\alpha$ at fixed orbital radius $r=10M$; (b) the spin parameter $a/M$ at $r=10M$; (c) the orbital radius $r/M$ for a rapidly rotating black hole with spin $a=0.9M$.}\label{omegaper}
\end{figure}

Fig.~\ref{omegaper} illustrates the periastron precession frequency $\Omega_{\text{per}}$ as a function of the MOG parameter $\alpha$, spin parameter $a$, and orbital radius $r$. Fig.~\ref{alphaper} shows that $\Omega_{\text{per}}$ increases with the MOG parameter $\alpha$, particularly in the low-spin regime, indicating that MOG amplifies the radial curvature responsible for the non-closure of bound orbits. However, the relative contribution of MOG diminishes as the black hole spin increases, due to the growing influence of frame dragging effects. In Fig.~\ref{aper}, $\Omega_{\text{per}}$ is shown to decrease with spin parameter $a$, with the rate of decline being slightly more pronounced for larger $\alpha$. This trend is governed by a competition between two effects: gravitational curvature, which promotes periastron advance, and frame dragging, which tends to suppress it. As a result, for fixed $\alpha$, the frame dragging contribution enters with a negative sign (see Eq.(\ref{per_expansion})) and reduces the overall precession frequency. Fig.~\ref{rper} displays the radial decay of $\Omega_{\text{per}}$, with larger $\alpha$ values producing a steeper decline. This highlights the fact that MOG-induced modifications to the gravitational potential are localized in the strong-field region and become negligible at large radii, where the spacetime curvature approaches the Newtonian limit.

Together, these trends demonstrate that $\Omega_{\text{per}}$ increases with $\alpha$, decreases with spin, and drops rapidly with orbital radius. This behavior reflects how MOG modifies the radial structure of spacetime and offers a clean channel to test strong-field curvature effects. As with nodal precession, precise measurements of $\Omega_{\text{per}}$ in mildly eccentric orbits may provide empirical constraints on modified gravity.

The distinct dependencies of nodal and periastron precession on frame dragging and curvature, respectively, underscore their combined utility in probing the structure of spacetime. To further complement this picture, we now turn to the precession of a spinning gyroscope in Kerr-MOG spacetime---a direct test of gravitomagnetic coupling and rotational geometry.

\section{Spin Precession of a Gyroscope in Kerr-MOG Spacetime}
\label{LTprecession}

Having investigated orbital precessions due to spacetime geometry, we now turn to the behavior of local inertial frames. In particular, we examine the spin precession of a test gyroscope stationary with respect to a zero-angular-momentum observer (ZAMO) in Kerr-MOG spacetime. This effect provides a complementary probe of spacetime rotation, the gravitomagnetic structure of modified gravity, independently of geodesic motion.

In GR, the Lense-Thirring effect describes how rotating masses, such as Earth or black holes, drag local inertial frames \cite{Lense:1918zz}, resulting in a measurable precession of the gyroscope’s spin axis. This effect reflects how the rotation of massive objects distorts the surrounding spacetime, leading to measurable deviations in the motion of test bodies. Gravity Probe B provided direct experimental confirmation of this phenomenon in Earth’s gravitational field \cite{Everitt:2011hp}. In the strong-field regime of rotating black holes, such frame dragging effects become significantly amplified. Importantly, for non-equatorial spherical orbits, the precession of a gyroscope is sensitive not only to the black hole spin but also to the geometry off the equatorial plane, introducing an explicit dependence on the orbital tilt angle $\zeta$.

In Kerr-MOG spacetime, where the gravitational interaction is enhanced by the MOG parameter $\alpha$, the structure of spacetime---and thus the spin precession---may differ significantly from that in standard Kerr geometry. The Lense–Thirring precession frequency of a gyroscope at rest in a general stationary axisymmetric spacetime is given by \cite{Chakraborty:2013naa,Straumann:2013spu}:
\begin{eqnarray}
	\Omega_{\texttt{LT}}=\frac{1}{2\sqrt{-g}}\bigg(\sqrt{g_{\theta \theta}}\Big(g_{t\phi,r}-\frac{g_{t\phi}}{g_{tt}}g_{tt,r}\Big) \hat{\theta}-\sqrt{g_{rr}}\Big(g_{t\phi,\theta}-\frac{g_{t\phi}}{g_{tt}} g_{tt,\theta}\Big)\hat{r}\bigg),
\end{eqnarray}
where $g$ is the determinant of the metric. The precession frequency arises from the spatial derivatives of the off-diagonal component $g_{t\phi}$, which governs the frame dragging, and the lapse function $g_{tt}$, associated with gravitational redshift. The radial and polar derivatives encode how these quantities vary across spacetime, with the $\hat{\theta}$-component dominating equatorial frame dragging and the $\hat{r}$-component capturing axial twisting near the rotation axis. The overall magnitude of $\Omega_{\texttt{LT}}$ thus encodes the local shear of frame dragging, which is enhanced in MOG due to the stronger curvature associated with the nonzero value of $\alpha$.

\begin{figure}
\begin{center}
\subfigure[]{\label{rlt5a}
\includegraphics[width=0.3\textwidth]{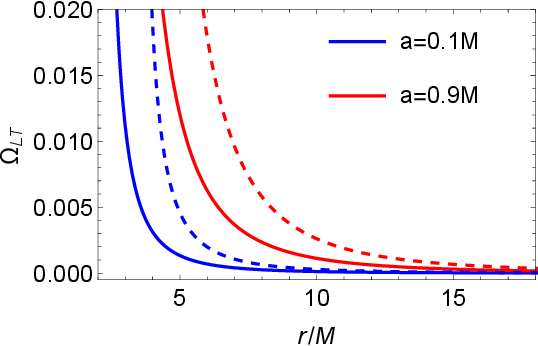}}
\hspace{1cm}
\subfigure[]{\label{rlt5b}
\includegraphics[width=0.3\textwidth]{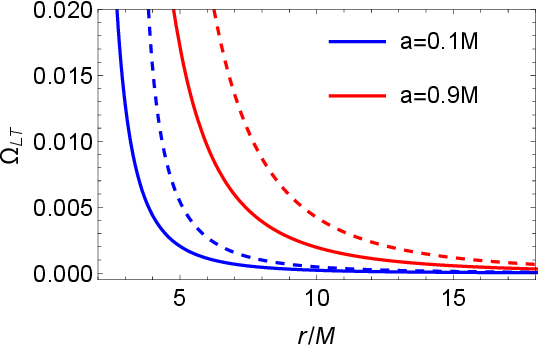}}
\caption{Lense–Thirring precession frequency $\Omega_{\texttt{LT}}$ as a function of the orbital radius. (a) Equatorial orbits ($\zeta = 0$). (b) Tilted orbits $\zeta = \frac{4\pi}{9}$. Solid curves: $\alpha = 0$ (Kerr); dashed curves: $\alpha = 1$ (Kerr-MOG).}\label{rlt}
\end{center}
\end{figure}

We first examine the radial dependence of $\Omega_{\texttt{LT}}$ for both equatorial ($\zeta = 0$) and tilted orbits ($\zeta = \frac{4\pi}{9}$), as shown in Fig.~\ref{rlt}. In both panels, $\Omega_{\texttt{LT}}$ decreases monotonically with radius and vanishes asymptotically, consistent with the characteristic $r^{-3}$ decay of the frame dragging in the weak-field limit. At fixed $r$, the precession frequency increases with both $a$ and $\alpha$, reflecting the combined effect of spin and MOG-induced enhancement of gravitomagnetic curvature. Importantly, the precession is significantly stronger in tilted orbits (Fig.~\ref{rlt5b}) than in the equatorial case (Fig.~\ref{rlt5a}), especially in the strong-field regime ($r \lesssim 10M$). This amplification arises because the misalignment between the spin axis and orbital plane introduces a stronger projection of the gyroscope’s spin onto the frame dragging direction, increasing the precession rate. Moreover, the sensitivity to $\alpha$ becomes more pronounced at small radii and high inclinations, where the nonlinear effects of the modified gravity theory accumulate. These features suggest that gyroscopic precession could serve as a clean observational window into the frame dragging structure of alternative gravity theories.

\begin{figure}
\begin{center}
\subfigure[]{\label{zetalt6a}
\includegraphics[width=0.3\textwidth]{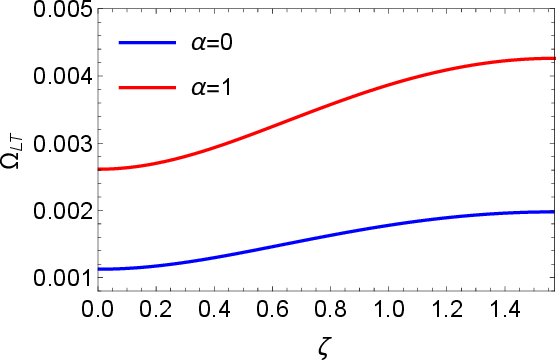}}
\quad
\subfigure[]{\label{alt6b}
\includegraphics[width=0.305\textwidth]{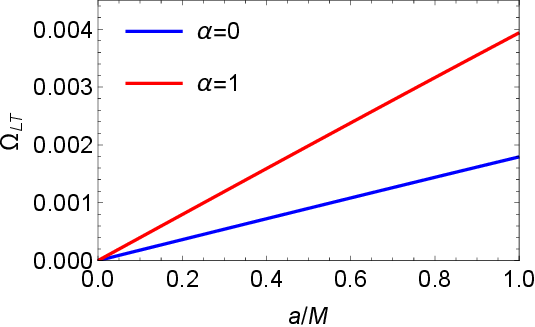}}
\quad
\subfigure[]{\label{alphalt6c}
\includegraphics[width=0.305\textwidth]{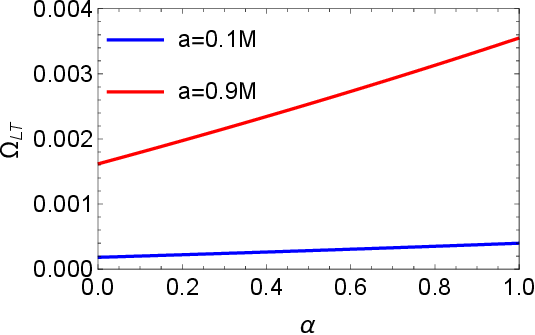}}
\caption{Lense-Thirring precession frequency $\Omega_{\texttt{LT}}$ as a function of: (a) tilt angle $\zeta$ for $a = 0.9M$; (b) spin parameter $a/M$ for $\zeta = \frac{\pi}{4}$; (c) MOG parameter $\alpha$ for $\zeta = \frac{\pi}{4}$. In all cases, the orbital radius is fixed at $r = 10M$.}\label{omegalt}
\end{center}
\end{figure}

To further disentangle the role of key parameters, we plot $\Omega_{\texttt{LT}}$ as functions of the tilt angle $\zeta$, spin $a$, and MOG parameter $\alpha$ at fixed radius $r = 10M$ in Fig.~\ref{omegalt}. Fig.~\ref{zetalt6a} shows that $\Omega_{\texttt{LT}}$ increases monotonically with the tilt angle $\zeta$. As the inclination grows, the projection of the gyroscope’s spin onto the direction of spacetime rotation becomes more pronounced, thereby enhancing the observed precession. Even in the equatorial plane ($\zeta = 0$), the precession remains finite. This reflects the fact that gyroscopic precession arises from the parallel transport of the spin vector in curved spacetime: the presence of frame dragging, governed by the off-diagonal metric component $g_{t\phi}$, induces a twist of local inertial frames, resulting in nonzero $\Omega_{\texttt{LT}}$ even for equatorial observers.

In Fig.~\ref{alt6b}, we observe that $\Omega_{\texttt{LT}}$ increases steeply with the black hole spin $a$, vanishing in the Schwarzschild limit ($a = 0$), where frame dragging is absent. This confirms that the precession originates from spacetime rotation. The steeper rise of $\Omega_{\texttt{LT}}$ at larger $\alpha$ reflects the enhancement of the spin–curvature coupling in MOG: the increased gravitational strength amplifies the twisting of inertial frames. This result highlights how spin and MOG-induced modifications together amplify gyroscopic precession, especially in the strong-field region, making $\Omega_{\texttt{LT}}$ a sensitive probe of both rotation and deviations from GR.

Fig.~\ref{alphalt6c} further illustrates that $\Omega_{\texttt{LT}}$ grows monotonically with the MOG parameter $\alpha$, at fixed tilt angle and radius. The rate of increase becomes more pronounced for higher black hole spin. This behavior reflects how the enhanced gravitational coupling in MOG strengthens the local curvature gradients that govern frame dragging. As a result, the gyroscope experiences a stronger gravitomagnetic torque, which increases the precession rate. These findings reinforce the idea that $\Omega_{\texttt{LT}}$ offers a clean and sensitive probe of spacetime rotation and scalar modifications in strong-field gravity.

In summary, gyroscopic spin precession in Kerr-MOG spacetime exhibits clear and distinct dependencies on orbital radius, tilt, black hole spin, and the MOG parameter $\alpha$. It encodes vital information about the rotational and curvature structure of the background, and provides a clean, local measure of frame dragging, complementary to trajectory-based precession diagnostics. The strong $\alpha$-sensitivity of $\Omega_{\texttt{LT}}$, especially at large inclinations, suggests that spin precession may serve as a robust observational signature of modified gravity.

\section{Summary}
\label{summary}

In this work, we have presented a comprehensive study of the spherical orbital dynamics of massive particles in Kerr-MOG spacetime within the STVG framework. Spherical orbits---characterized by constant radial coordinate but nontrivial polar motion---exhibit richer dynamics than equatorial orbits, as they depend sensitively on both the orbital tilt angle $\zeta$ and the MOG parameter $\alpha$. These parameters significantly modify the conserved quantities and relativistic precession effects, offering new opportunities to probe deviations from GR in the strong-field regime.

We began by reviewing the Kerr-MOG black hole solution and its key departures from the standard Kerr geometry, including the enhanced gravitational coupling arising from the additional scalar and vector fields. Using the Hamilton–Jacobi formalism, we derived the geodesic equations and systematically analyzed the conserved energy, angular momentum, and the Carter constant associated with spherical orbits. Our numerical results reveal that increasing $\alpha$ shifts the ISSO outward and strengthens the gravitational binding, while increasing the orbital tilt angle enhances the Carter constant and alters orbital stability. These trends highlight the intricate interplay between orbital geometry and the underlying gravitational field structure in MOG.

We then explored two fundamental types of relativistic orbital precession: nodal precession, arising from frame dragging, and periastron precession, governed by spacetime curvature. We demonstrated that both precession frequencies are highly sensitive to the black hole spin $a$, the MOG parameter $\alpha$, and the orbital radius $r$. The nodal precession frequency increases monotonically with both $a$ and $\alpha$, reflecting the cumulative effects of rotation and enhanced gravitomagnetic curvature. In contrast, periastron precession exhibits a more intricate dependence, with contributions from both curvature-induced orbit non-closure and spin-induced frame dragging. MOG amplifies the curvature effect, while black hole spin can partially counteract it, underscoring the need for simultaneous consideration of both phenomena in precise measurements.

In addition to orbital precessions, we analyzed the Lense-Thirring spin precession of a test gyroscope stationary relative to a ZAMO. Our results show that the precession frequency $\Omega_{\texttt{LT}}$ grows with both the spin and MOG correction, and is further amplified for inclined orbits due to enhanced coupling between the gyroscopic spin and the gravitomagnetic field. Notably, we found that even in the equatorial configurations, $\Omega_{\texttt{LT}}$ remains finite due to the intrinsic twisting of inertial frames caused by frame dragging. This analysis demonstrates that gyroscopic spin precession provides a clean, local measure of spacetime rotation---distinct from trajectory-based diagnostics---and is particularly sensitive to MOG-induced modifications in the near-horizon region.

Our findings suggest that spherical orbits, together with their associated precession phenomena, encode rich physical information about the interplay between spacetime curvature, rotation, and possible deviations from GR. These effects are especially relevant for astrophysical scenarios involving tilted accretion disks, jet precession in active galactic nuclei (such as M87*), and compact binary systems with inclined orbital configurations. In particular, the sensitivity of nodal and spin precession to the MOG parameter offers a potential observational window to distinguish MOG from GR using future VLBI or gravitational wave observations.

While this work focuses on Kerr-MOG black holes within STVG, it would be valuable to extend this analysis to other modified gravity theories, such as Einstein-dilaton-Gauss-Bonnet gravity or dynamical Chern-Simons gravity. Comparative studies could help identify distinctive signatures across competing models, thereby improving prospects for testing the foundations of gravity in the strong-field regime.

Future directions include extending our analysis to eccentric and inspiraling orbits, integrating relativistic precession effects into gravitational waveform modeling, and assessing their detectability with next-generation experiments such as the EHT, LISA, and pulsar timing arrays. Such efforts could pave the way toward robust, multi-messenger tests of GR and alternative theories of gravity in the coming decade.

\section*{Acknowledgements}
This work was supported by the National Natural Science Foundation of China (Grants No. 12222302), and National Key Research and Development Program of China (No. 2024YFC2207400).

\end{document}